\documentclass[aps,prapplied,reprint,floatfix,citeautoscript,longbibliography,superscriptaddress]{revtex4-2}
\usepackage{graphicx}
\usepackage{bm,amsmath,amssymb,mathrsfs,dcolumn}
\usepackage[colorlinks=true, linkcolor=blue, citecolor=blue, urlcolor=blue, linktoc=page, bookmarks=false, pdfstartview={FitH}, pdfborder={0 0 0.0 [3 3]}]{hyperref} % HyperLink
\usepackage[usenames]{xcolor}
\hypersetup{pdfborder=0 0 0,colorlinks=true,citecolor=blue,linkcolor=blue}

\begin{document}
\title{Anticipative Tracking with the Short-Term Synaptic Plasticity of Spintronic Devices}
\author{Qi Zheng}
\affiliation{Center for Advanced Quantum Studies and Department of Physics, Beijing Normal University, Beijing 100875, China}
\affiliation{Center for Quantum Computing, Peng Cheng Laboratory, Shenzhen 518005, China}
\author{Yuanyuan Mi}
\affiliation{Center for Neurointelligence, Chongqing University, Chongqing 400044, China}
\affiliation{Center for Artificial Intelligence, Peng Cheng Laboratory, Shenzhen 518005, China}
\author{Xiaorui Zhu}
\affiliation{Center for Advanced Quantum Studies and Department of Physics, Beijing Normal University, Beijing 100875, China}
\affiliation{Center for Quantum Computing, Peng Cheng Laboratory, Shenzhen 518005, China}
\author{Zhe Yuan}
\email{zyuan@bnu.edu.cn}
\affiliation{Center for Advanced Quantum Studies and Department of Physics, Beijing Normal University, Beijing 100875, China}
\affiliation{Center for Quantum Computing, Peng Cheng Laboratory, Shenzhen 518005, China}
\author{Ke Xia}
\affiliation{Center for Advanced Quantum Studies and Department of Physics, Beijing Normal University, Beijing 100875, China}
\affiliation{Center for Quantum Computing, Peng Cheng Laboratory, Shenzhen 518005, China}
\affiliation{Beijing Computational Science Research Center, 100193 Beijing, China}
\affiliation{Shenzhen Institute for Quantum Science and Engineering and Department of Physics, Southern University of Science and Technology, Shenzhen 518055, China}
\date{\today}
\begin{abstract}
Real-time tracking of high-speed objects in cognitive tasks is challenging in the present artificial intelligence techniques because the data processing and computation are time-consuming resulting in impeditive time delays. A brain-inspired continuous attractor neural network (CANN) can be used to track quickly moving targets, where the time delays are intrinsically compensated if the dynamical synapses in the network have the short-term plasticity. Here, we show that synapses with short-term depression can be realized by a magnetic tunnel junction, which perfectly reproduces the dynamics of the synaptic weight in a widely applied mathematical model. Then, these dynamical synapses are incorporated into one-dimensional and two-dimensional CANNs, which are demonstrated to have the ability to predict a moving object via micromagnetic simulations. This portable spintronics-based hardware for neuromorphic computing needs no training and is therefore very promising for the tracking technology for moving targets.
\end{abstract}
\maketitle

%%%%%%%%10%%%%%%%%20%%%%%%%%30%%%%%%%%40%%%%%%%%50%%%%%%%%60%%%%%%%%70%%%%%%%%80
\section{Introduction}

Computations using artificial neural networks have shown tremendous efficiency in applications, such as pattern recognition \cite{Jain02,Kaiming16} and natural language programming \cite{Tom18}. These computations usually require a finite processing time and hence bring challenges to those tasks involving a time limit, e.g., tracking objects that are quickly moving. Object tracking has been performed with various algorithms including the correlation filter \cite{KCF} and artificial neural networks \cite{RNN}. Visual object tracking is a basic cognitive ability of animals and human beings. Some particular mechanisms are intrinsically adopted in biological brains \cite{Blair95,Nijhawan08} to compensate the finite processing time in neural systems. A bio-inspired algorithm is developed to incorporate the delay compensation into a tracking scheme and allow it to predict fast moving objects. It has been implemented based on a special neural network called a continuous attractor neural network (CANN) \cite{CANN1}, which has been experimentally observed in animals' brains \cite{Kim2017,HD}. Anticipative tracking can be achieved using CANNs with short-term depression (STD) of synaptic efficacy \cite{CANN delay}, spike frequency adaptation of neurons \cite{Mi14} or negative feedback from a neighboring layer \cite{Zhang12}.

STD is one of the typical properties of short-term synaptic plasticity and naturally exists in biological neural systems \cite{Nichollsbook}. Because of the depletion of the neurotransmitter at the end of the presynaptic neuron consumed during spikes, the synaptic efficacy becomes weaker than its normal amplitude shortly after spikes and recovers with a time constant $\tau$ \cite{dynamic1,dynamic2}. This special property of synapses intrinsically introduces a negative feedback into a CANN, which therefore sustains spontaneous traveling waves. If the CANN with negative feedback is driven by a continuously moving input, the resulting network state can lead the external drive at an intrinsic speed of traveling waves larger than that of the external input. Anticipative tracking is therefore achieved \cite{CANN delay}.

The recently developed Tianjic chips have integrated many CANNs into their architecture and show the ability to track objects in the application \cite{Tianjic}. Unfortunately, there are no dynamical synapses with short-term plasticity; thus, predicting the trajectory of a moving object is not yet possible. Another application scenario is the video captured by the high-speed cameras, which can be up to a million frames per second \cite{DVS}. Therefore, the real-time tracking of an object in the high-speed video requires a very quick response in devices and a dynamical synapse with controllable STD is highly desirable. The short-term synaptic plasticity has also been discovered in resistive, ferroelectric and even 2D van der Waals layered materials \cite{Zhu18,Gechen18,Gechen20,Ailbart10,Ohno11,Kim18,Seo19}, which are, in principle, potential candidates for developing CANN hardware to perform tracking tasks. The STD in these materials is usually related to the process of atomic diffusion. Unlike the diffusion mechanism, the STD time scale of spintronic devices, e.g. magnetic tunnel junctions (MTJs), is determined by the dissipation of spin angular momentum in magnetization dynamics, which is in the gigahertz regime resulting in a small time constant $\tau$. Moreover, this time scale can be tuned by manipulating the damping parameters via material components \cite{ZhaoYawen2018} and magnetic configuration \cite{YuanPRL2014}. This flexibility makes MTJs easier to be applied in the CANN for tracking tasks than other materials. The nonvolatile memory and large endurance ($>10^{15}$) of spintronic devices make them highly suitable for the hardware implementation of artificial neural networks \cite{Grollier16,Zhang2020}. Such spintronics-based portable devices with low energy consumption would have great potentials for applications. For instance, these devices can be embedded in a mobile equipment. 

In this article, we use the magnetization dynamics of MTJs to realize short-term synaptic plasticity. These dynamical synapses are then plugged into a CANN to achieve anticipative tracking, which is illustrated by micromagnetic simulations. As a proof of concept, we first demonstrate a prediction for a moving signal inside a one-dimensional (1D) ring-like CANN with 20 neurons. The phase space of the network parameters is discussed. Then, we consider a two-dimensional (2D) CANN with arrays of MTJs, which can be used to analyze moving objects in a video.

%%%%%%%%10%%%%%%%%20%%%%%%%%30%%%%%%%%40%%%%%%%%50%%%%%%%%60%%%%%%%%70%%%%%%%%80
\section{The Structure of a CANN}

\begin{figure}[t]
\includegraphics[width=0.8\columnwidth]{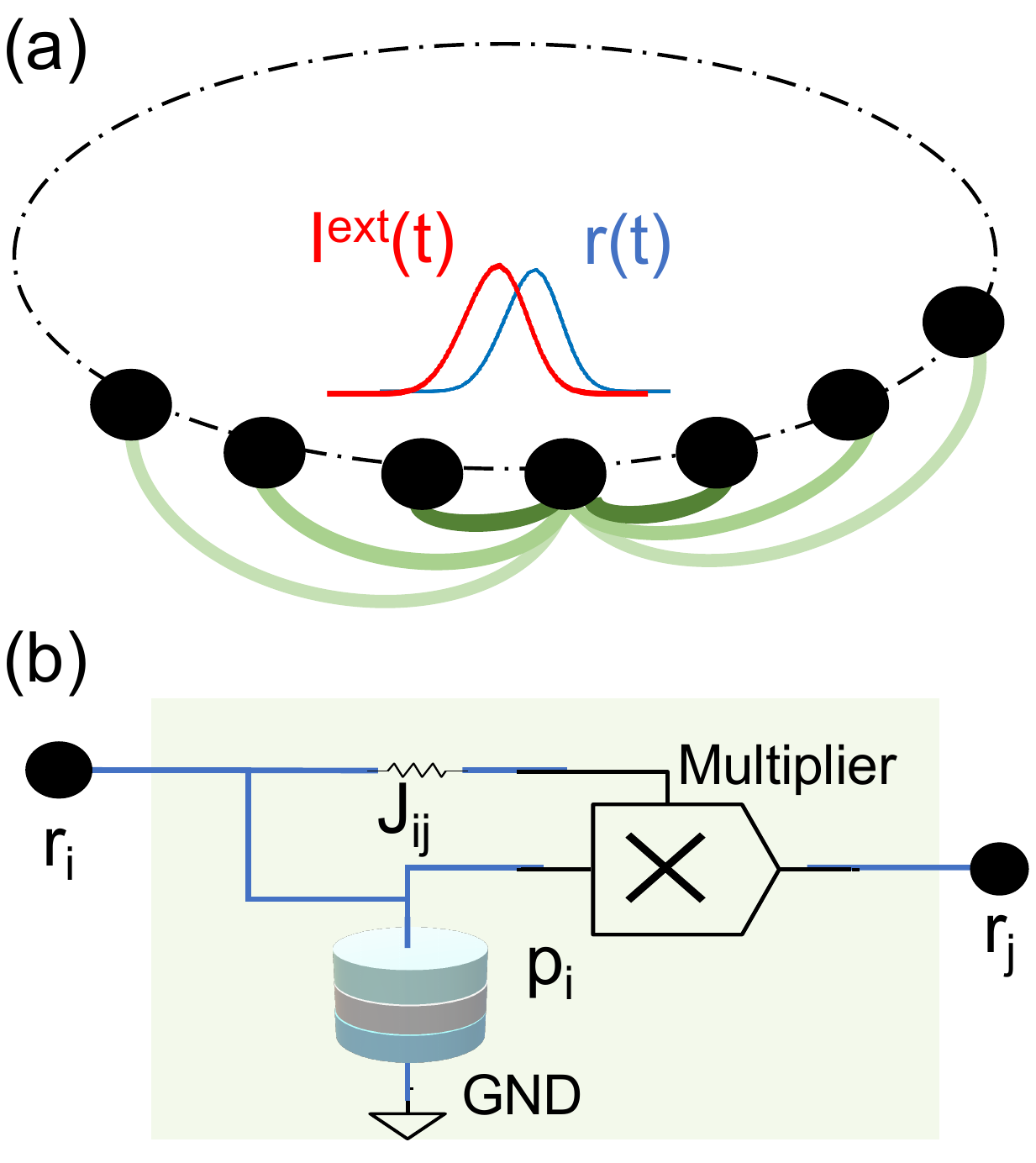}
\caption{(a) Schematic diagram of the 1D CANN. The black dots represent the neurons located at position $\mathbf x_i=(\cos\theta_i,\sin\theta_i)$. When external stimuli $I^{\rm ext}(t)$ move along the chain of neurons, the firing rate $r(t)$ of neurons is determined by the external input and the signal received from other neurons. The light-green line is the unidirectional connection, which allows the output of the $i$th neuron to be transmitted to the $j$th neuron, as indicated by Eq.~\eqref{eq:input}. (b) The hardware implementation of a synapse consisting of a constant component and a dynamical one with STD. The constant weight $J_{ij}$ is modeled by an electric resistor, and the dynamical synapse $p_i$ is simulated by an MTJ.}\label{fig:1}
\end{figure}

A CANN is a special type of recurrent neural network that has translational invariance. We first use a 1D model as an example to illustrate the structure and functionality of a CANN. As shown in Fig.~\ref{fig:1}(a), a number of neurons are connected to form a closed chain. We use an angular coordinate $\theta_i$ to describe the neuron positions $\mathbf x_i=(\cos\theta_i,\sin\theta_i)$, which are uniformly distributed on the unit circle. The $i$th neuron follows the dynamical equation \cite{CANN delay}
\begin{equation}
\tau_s \dot U_i (t) =-U_i(t)+I_i(t),\label{eq:neuron}
\end{equation}
where $U_i$ denotes the population-averaged synaptic current to the $i$th neuron. The second term on the right-hand side $I_i$ is the total input, including the external stimuli $I_i^{\rm ext}$ and the signals transmitted from other neurons, i.e.,
\begin{equation}
I_i(t)=I_i^{\rm ext}(t)+\sum_j J_{ji}p_j(t)r_j(t).\label{eq:input}
\end{equation}
The external input has a Gaussian profile, and its center moves inside the network. To avoid the divergence of the network, it is essential to have a global inhibition; here, we consider a normalization of $U_i$ for convenience in the implementation. Specifically, we define
\begin{equation}
u_i(t)=\frac{U_i(t)-\min[U_i(t)]}{\max[U_i(t)]-\min[U_i(t)]},
\end{equation}
and the firing rate of the $i$th neuron is given by $r_i(t)\equiv u_i^2(t)/k$ in Eq.~\eqref{eq:input}. Here, the parameter $k$ denotes the inhibition strength.

It is worth noting that we focus on synapses in this work and do not consider the particular hardware implementation of the neuron. This is because a CANN without dynamical synapses was realized based on a mixed digital and analogue silicon circuit \cite{CANN2000} and in the Tianjic chip \cite{Tianjic}. Moreover, as we have demonstrated in an independent work \cite{Zheng20}, Eq.~\eqref{eq:neuron} indicates a decayed dynamics, and this neuron can be replaced by a single MTJ. The precessional period of the MTJ determines the parameter $\tau_s$ in Eq.~\eqref{eq:neuron}, which is usually required to be two orders of magnitude smaller than the STD time scale $\tau$. We take the limit of $\tau_s\rightarrow0$ in this work for simplicity.

The key characteristic of the CANN that we propose is the dynamical synapses; every synapse connects a pair of neurons, as illustrated by the green lines in Fig.~\ref{fig:1}(a). In Eq.~\eqref{eq:input}, the synaptic efficacy contains two factors, i.e., a constant weight $J_{ji}$ that is only a function of the distance of the $i$th and $j$th neurons and a dynamical weight $p_j$ that depends on the neuron dynamics in the recent past. To implement the synapses, we propose the structure sketched in Fig.~\ref{fig:1}(b), where the constant weight $J_{ij}$ corresponds to a constant electric resistor. Following the standard model in computational neurosciences, we adopt a dimensionless, normalized value
\begin{equation}
J_{ji}=\frac{b}{a}\exp\left[-\frac{(\mathbf x_i-\mathbf x_j)^2}{2a^2}\right],\label{eq:const}
\end{equation}
with $b$ and $a$ being the parameters for controlling the strength and range of the synaptic connections, respectively. The dynamical synapses with STD can be realized by MTJs, and the driving current density injected into the MTJ depends on the firing rate of the neuron. The specific definition of its efficacy will be illustrated below in Eq.~\eqref{eq:efficacy}. In the end, the signals transmitted through the electric resistor and through the MTJ are multiplied as the input to the next neuron.

The eventual performance of the CANN is examined by comparing the instantaneous distributions of the external stimuli $I_i^{\rm ext}(t)$ and the firing rate $r_i(t)$. Both $I_i^{\rm ext}(t)$ and $r_i(t)$ form bump-like distributions and move in the CANN. If the center of $r_i(t)$ moves ahead of the external stimuli, anticipative tracking is achieved. Otherwise, one has delayed tracking.

%%%%%%%%10%%%%%%%%20%%%%%%%%30%%%%%%%%40%%%%%%%%50%%%%%%%%60%%%%%%%%70%%%%%%%%80

\section{Short-term Plasticity of MTJs}

The distinct feature of a dynamical synapse with STD is the temporarily reduced efficacy right after firing of the associated neuron, which can be gradually recovered over a longer time scale. This dynamical behavior can be found in an MTJ consisting of two thin ferromagnetic layers separated by an insulator. One of the ferromagnetic layers has a fixed magnetization, which is usually pinned by a neighboring antiferromagnetic material via the so-called exchange bias. The magnetization of the other (free) layer can be excited to precess by an electric current via the spin-transfer torque. The precession will not stop immediately after the end of the injected current but will gradually decay due to Gilbert damping. The electrical resistance of the MTJ, which depends on the relative magnetization orientation of the two ferromagnetic layers, therefore exhibits a temporary variation after the excitation. Then, we can implement the STD with the dynamics of MTJs by taking the injected electric current and electric resistance as the input and output, respectively.

\begin{figure}[t]
\includegraphics[width=\columnwidth]{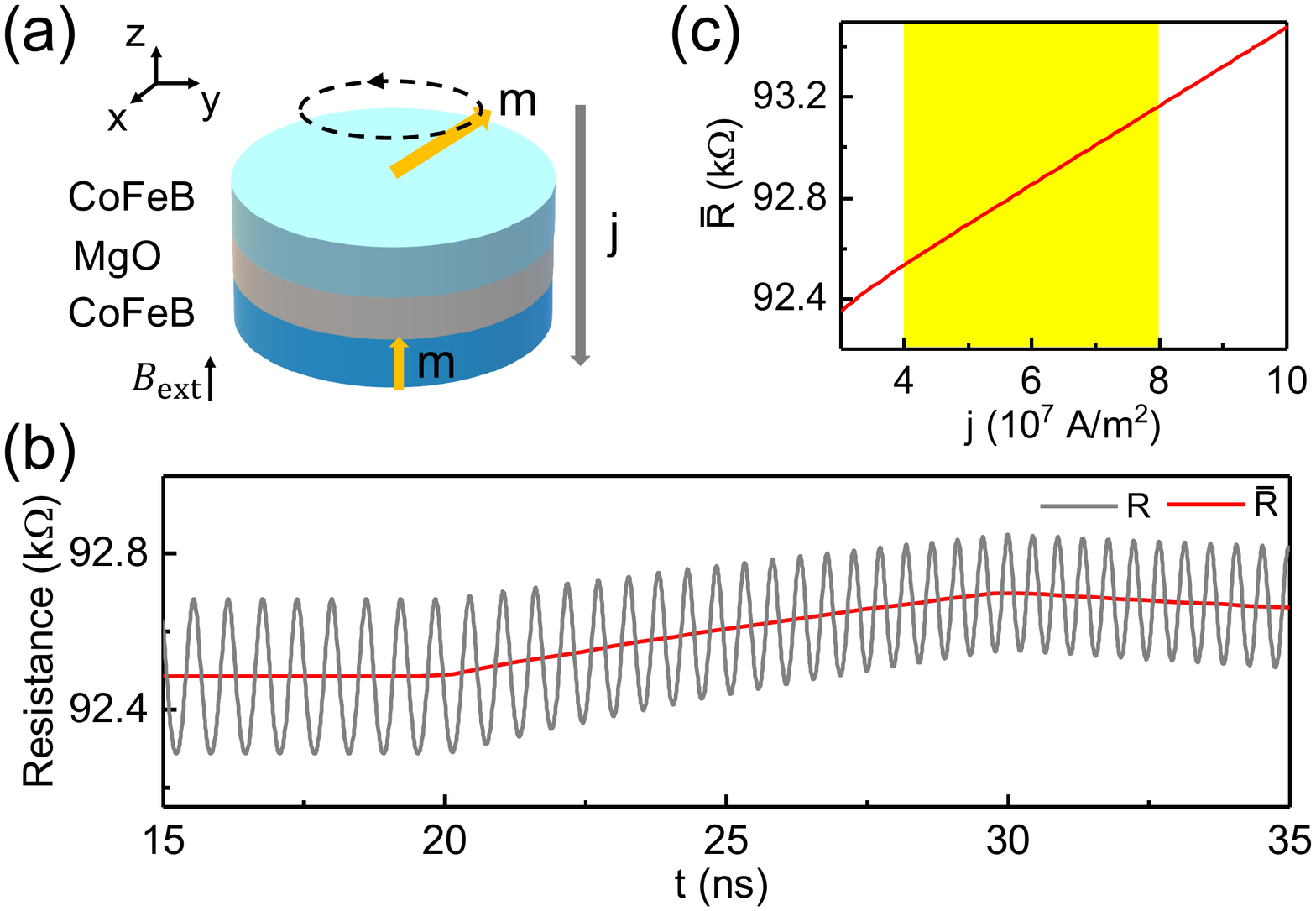}
\caption{(a) Schematic illustration of an MTJ consisting of a fixed CoFeB layer, insulating MgO layer and free CoFeB layer from bottom to top. The upward magnetization of the fixed layer is pinned by exchange bias from an antiferromagnetic substrate, and the free layer has an in-plane magnetic anisotropy, which precesses under an injected electric current. (b) Calculated electrical resistance of the MTJ as a function of time. The injected current density is $8\times10^7$~A/m$^2$ from $t=20$~ns to $30$~ns and $4\times10^7$~A/m$^2$ otherwise. The red line represents the average resistance $\bar R$ as a function of time. (c) Average resistance $\bar R$ as a function of the current density for stable oscillation. The yellow shadow region is the current density range used in this work.}\label{fig:2}
\end{figure}
To explicitly illustrate the functionality, we consider a cylindrical CoFeB/MgO/CoFeB MTJ with a radius of 20~nm \cite{Hong18}, as schematically shown in Fig.~\ref{fig:2}(a), where the free layer, insulating layer and fixed layer are stacked from top to bottom. The thicknesses of the free and insulating layers are 2~nm and 1.1~nm, respectively \cite{Hong18}. The fixed layer has a perpendicular magnetization along $+z$, and the free layer has a dominant in-plane anisotropy due to the demagnetization energy. For CoFeB, we choose the following material parameters: the saturation magnetization $M_s=10^6$~A/m, exchange stiffness $A=2\times10^{-11}$~J/m \cite{Zhao17}, and Gilbert damping $\alpha=3\times10^{-4}$ \cite{foot1}.

In the presence of an electric current density $j$ through the MTJ, the magnetization of the free layer is excited to precess following the generalized Landau-Lifshitz-Gilbert equation \cite{Slonczewski96,Berger96,Zhang02}.
\begin{eqnarray}
\dot{\mathbf m}&=&-\gamma\mathbf m\times\mathbf H_{\rm eff}+\alpha\mathbf m\times\dot{\mathbf m}\nonumber\\
&&+\tau\epsilon\mathbf m\times\mathbf m_p\times\mathbf m-\beta\tau\mathbf m\times\mathbf m_p.\label{eq:llg}
\end{eqnarray}
Here, $\mathbf m$ is the local magnetization direction in the free layer, and $\mathbf H_{\rm eff}$ is the effective magnetic field, including the exchange, anisotropy and demagnetization fields, as well as an external magnetic field $B_{\rm ext}=0.5$~T along the $+z$ axis to stabilize the precession. The last two terms in Eq.~\eqref{eq:llg} are the adiabatic and nonadiabatic spin-transfer torques, respectively, where $\mathbf m_p$ denotes the magnetization direction of the fixed magnetic layer and the magnitude of the torque $\tau=(\gamma\hbar P/\mu_0 eM_s t)j$ depends on the current polarization $P$, the current density $j$, the saturation magnetization $M_s$ and the free-layer thickness $t$. The Slonczewski parameter $\epsilon=\Lambda^2/[(\Lambda^2+1)+(\Lambda^2-1)\mathbf m\cdot\mathbf m_p]$ characterizes the angular dependence of the torque with the dimensionless parameter $\Lambda=1.5$ used in our simulations. $\beta$ is the nonadiabaticity of the spin-transfer torque and is usually much smaller than one. The dynamic equation~\eqref{eq:llg} is solved numerically using the micromagnetic simulation program MuMax3 \cite{Mumax3}, and the free magnetic layer is discretized into a $10\times10\times1$ grid.

We apply the Julliere formula to estimate the tunneling magnetoresistance ratio
\begin{equation}
\mathrm{TMR}=\frac{2P_{\rm free} P_{\rm fixed}}{1-P_{\rm free} P_{\rm fixed}},
\end{equation}
where $P_{\rm free}$ and $P_{\rm fixed}$, representing the spin polarization of the free and fixed layers, respectively, are both set to 0.6 for CoFeB. The resistance $R$ of the MTJ is determined by the instantaneous magnetization $\mathbf m$ of the free layer \cite{Julliere75}:
\begin{equation}
R(\mathbf m)=\frac{R_{\rm P}}{2}\left[1+\mathrm{TMR}-\mathrm{TMR}(\mathbf m\cdot\mathbf m_p)\right],
\end{equation}
and the resistance of the MTJ for the parallel configuration $R_{\rm P}=71.6~\mathrm k \Omega $ \cite{Zhu16}.

Note that $R$ depends not only on the instantaneous current density but also on its historically recent dynamics. As plotted in Fig.~\ref{fig:2}(b), a steady oscillatory state is found for a current density $4\times10^7$~A/m$^2$ at $t\le20$~ns. Then, the current density is increased to $8\times10^7$~A/m$^2$ for the next 10~ns and returns to $4\times10^7$~A/m$^2$ afterwards. It can be seen that the electric resistance of the MTJ gradually increases or decreases in an oscillatory way due to the change in $j$ before reaching the corresponding steady state. In this work, the high-frequency oscillations are artificially filtered and we take the average resistance $\bar R$ as the weight of the dynamical synapses. In real devices, this filtering can be easily done by a standard Butterworth low pass filter. $\bar R$ at the steady states is plotted in Fig.~\ref{fig:2}(c) as a function of input current density $j$, and we restrict $j$ in this work to the range from $j_{\min}=4\times10^7$~A/m$^2$ to $j_{\max}=8\times10^7$~A/m$^2$, resulting in $\bar R_{\min}=92.5~\mathrm k\Omega$ and $\bar R_{\max}=93.1~\mathrm k\Omega$.

\begin{figure}[t]
\centering
\includegraphics[width=\columnwidth]{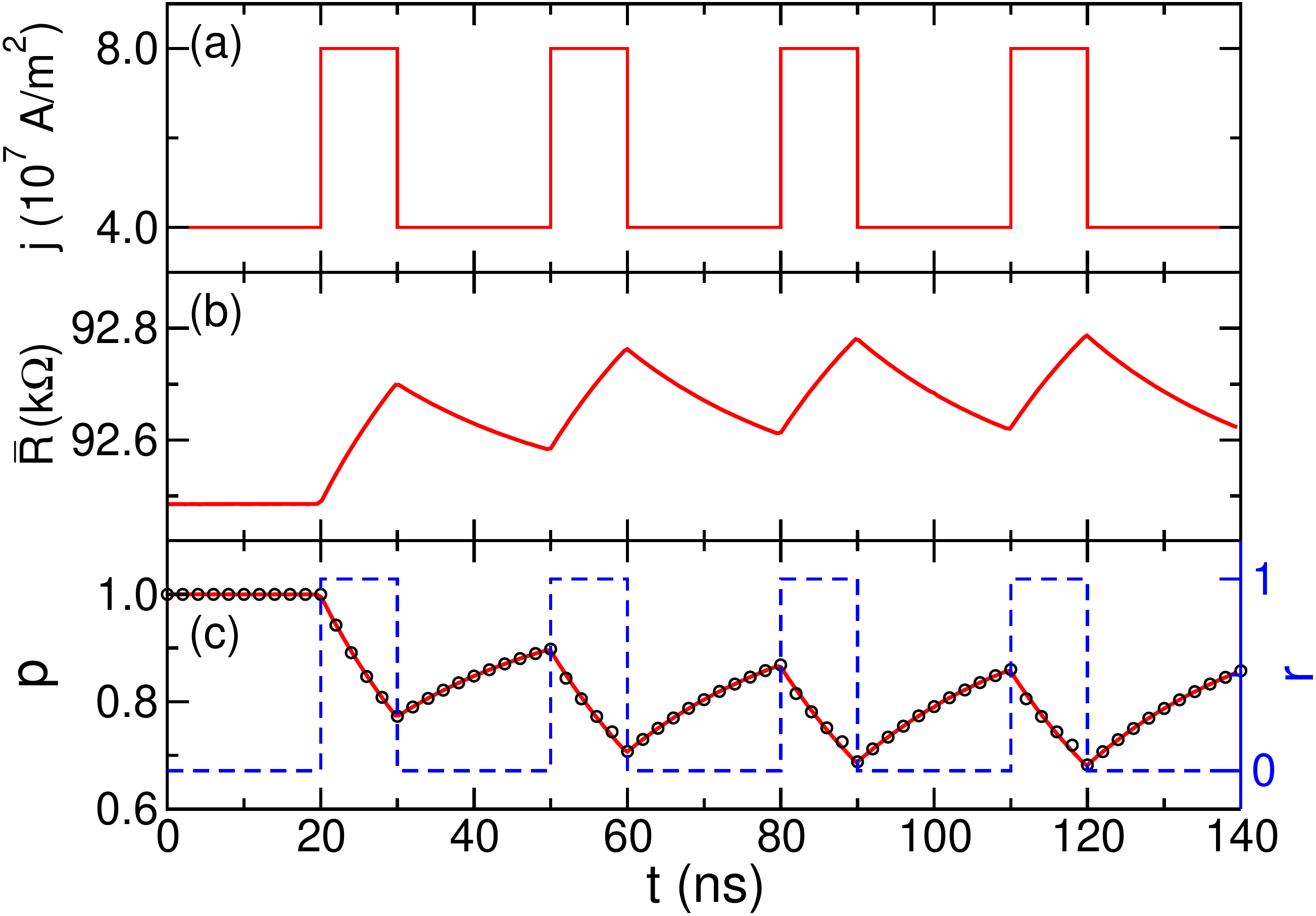}
\caption{(a) Current density injected into the MTJ as a function of time to drive the magnetization precession of the free layer. (b) Calculated average resistance $\bar R$ of the MTJ due to the precessional motion of the free-layer magnetization. At $t\le 20$~ns, the MTJ is at the steady oscillatory state with $j=4\times10^7$~A/m$^2$, and $\bar R$ is a constant. $\bar R$ increases and decreases gradually with increasing or decreasing $j$, indicating the short-term memory effect. (c) Synaptic efficacy $p$ obtained from the calculated $\bar R$ using Eq.~\eqref{eq:efficacy} as a function of time (red solid line). The black empty circles are the efficacy of a dynamic synapse, which is the solution of phenomenological Eq.~\eqref{eq:std}. The firing rate is denoted by the blue dashed line. The parameters $\tau=25$~ns and $\tilde{\eta}=0.79$ are used in Eq.~\eqref{eq:std}.}
\label{fig:3}
\end{figure}
To explicitly examine the STD behavior, we artificially apply a time-dependent current density, as shown in Fig.~\ref{fig:3}(a), to mimic a piecewise firing rate of a single neuron that has influence on the variable weight of the dynamic synapse. Then, the calculated $\bar R$ of the MTJ is plotted in Fig.~\ref{fig:3}(b), which shows the short-term memory effect. As the current density changes, the resistance gradually varies towards the value for the steady state. Then, we define a dimensionless synaptic efficacy that is associated with a specific neuron as
\begin{equation}
p(t)\equiv[\bar R_{\max}-\bar R(t)]/(\bar R_{\max}-\bar R_{\min})\label{eq:efficacy}
\end{equation}
and show $p$ in Fig.~\ref{fig:3}(c) as a function of time. When the current density changes, indicating a different firing rate, the synaptic efficacy $p$ exhibits an exponential variation corresponding to a short-term memory effect.

Usually, the dynamical synapse with STD is described by the following differential equation:
\begin{equation}
\tau \dot {\tilde p}(t)=1-\tilde p(t)-\tilde{\eta} \tilde p(t) r(t),\label{eq:std}
\end{equation}
where $\tau$ is the time scale of the STD and the parameter $\tilde{\eta}$ determines the strength of the STD in the synaptic efficacy. For the piecewise firing rate shown by the blue dashed line in Fig.~\ref{fig:3}(c), one can solve Eq.~\eqref{eq:std} analytically and obtain $p(t)$, plotted as the empty circles. Here, we take the parameters $\tau=25$~ns and $\tilde{\eta}=0.79$. The perfect agreement between the efficacy with the MTJ and the solution of Eq.~\eqref{eq:std} demonstrates that the dynamic synapses with STD can be realized very well using MTJs.

MTJs are promising candidates for synapses in spintronics-based neuromorphic computing, partly because the techniques for fabrication and integration are mature after being extensively studied as a memory device in the past two decades. The electrical and magnetic properties can be flexibly engineered including the in-plane or perpendicular magnetic anisotropy, ferromagnetic resonance frequency, the time scale of dynamical decay, electrical resistance and tunneling magnetoresistance, etc \cite{Khvalkovskiy2013,Chen2015}. Moreover, MTJs have other useful characteristics required as synapses in neuromorphic computing, such as the long-term potentiation \cite{Sengupta16}, multilevel memristors\cite{ZhangDM16}, and MTJs are even allowed to realize the spike timing dependent plasticity \cite{Srinivasan16}. All these properties significantly increase the feasibility of MTJs as integrated synapses in hardware of artificial neural networks. In the CANN to track moving objects, we utilize the the characteristic STD of MTJs, which is critical to compensate the computational delay in tracking. To realize the functionality, only a few percent of the STD strength is sufficient, as demonstrated in computational neuroscience models \cite{Fung2015}, because the intrinsic traveling waves in the network are very sensitive to the variation of synaptic weights. In addition, such variation can be magnified in electronic circuits. From the device point of view, this requires the signal-to-noise ratio is of the order of hundreds or larger.

%%%%%%%%10%%%%%%%%20%%%%%%%%30%%%%%%%%40%%%%%%%%50%%%%%%%%60%%%%%%%%70%%%%%%%%80
\section{Anticipative Tracking with Dynamical Synapses}
Since we have demonstrated that dynamic synapses with STD can be realized by MTJs, it is possible to implement hardware CANNs to perform the anticipative tracking of moving objects. In this section, two examples are illustrated to show the capability of CANNs with MTJs as the functional dynamic synapses. We first consider a 1D model system with twenty neurons forming a closed chain, as shown in Fig.~\ref{fig:1}(a), where a signal with a Gaussian distribution is moving around. In this model system, we show how anticipative tracking is achieved and discuss the influence of network parameters. Then, we consider a moving table tennis ball in a video and use a 2D CANN with arrays of MTJs to predict the motion of the ball.

\subsection{1D model}

\begin{figure}[t]
\includegraphics[width=0.95\columnwidth]{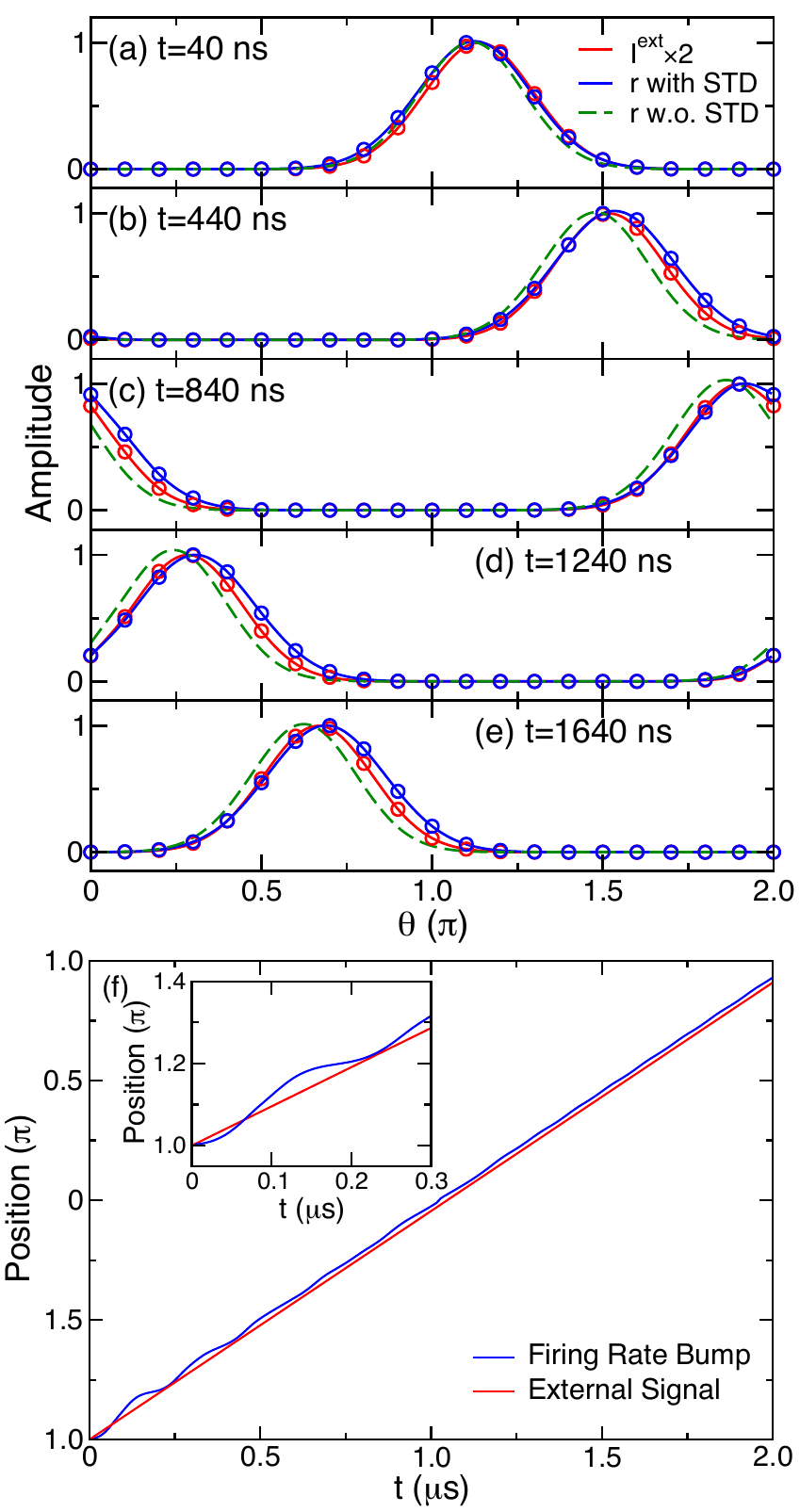}
\caption{(a)--(e) Anticipative tracking in the 1D CANN. The external signal (red solid line) with a Gaussian-type distribution moves from left to right. The firing rate of neurons is plotted as the blue solid line, which is ahead of the external signal at $t>100$~ns. The green dashed line represents the firing rate of the neurons without MTJs as the dynamic synapses for comparison. The green bump shows delayed tracking since it always follows the external signal with a finite delay in position. (f) The central positions of the external signal (red line) and the firing rate bump (blue line) as a function of time. Inset: the magnified plot at small $t$.}\label{fig:4}
\end{figure}

We first construct a 1D CANN consisting of twenty neurons that are uniformly distributed in the unit circle, as shown in Fig.~\ref{fig:1}(a). Then, the constant connection Eq.~\eqref{eq:const} is simplified as $J_{ji}=(b/a)\exp\{2\cos[(\theta_j-\theta_i)/2]/a^2\}$, with $a=0.5$ and $b=1.27$. The current density injected into the MTJs depends on the firing rate of the corresponding neuron $r_i$ and is explicitly computed by
\begin{equation}
j_i(t)=j_{\max}-\frac{2}{\pi}\arctan\left[\eta r_i(t)\right]\left(j_{\max}-j_{\min}\right).\label{eq:stdj}
\end{equation}
Here, $j_{\max}=2j_{\min}=8\times10^7$~A/m$^2$ is defined, and the parameter $\eta=0.8$ for controlling the STD strength. Then, we apply an external stimuli with a Gaussian distribution in the 1D CANN to simulate the received signal of a moving object:
\begin{equation}
I_i^{\rm ext}(t)=A \exp\left[-\frac{(\theta_i-\omega_{\rm ext} t)^2}{a^2}\right],
\end{equation}
where $A=0.5$ is the amplitude of the external pulse and $\omega_{\rm ext}=0.003$~rad/ns is the angular velocity of the external signal.

Owing to the STD effect, the CANN sustains traveling waves that are propagating in the network. As a consequence, the bump of the firing rate moves ahead of the external signal, indicating that anticipative tracking is achieved. Figure~\ref{fig:4} shows the firing rate of the 1D CANN as a response to a moving external signal. At the initial stage, e.g., $t=40$~ns, the firing rate of the neurons already forms a bump-like response to the external stimuli, and the firing rate (blue line) is slightly delayed compared with the right-going external signal (red line). At $t=440$~ns, the firing-rate distribution is already ahead of the external signal. Afterward, the firing rate is always in the lead and remains steady. The anticipative tracking can be clearly seen in Fig.~\ref{fig:4}(f), where the central positions of the firing rate and the external signal are plotted as a function of time. The central position of the external signal is a straight line due to the constant $\omega_{\rm ext}$. The firing rate exhibits a slight delay in the first 0.07~$\mu$s but catches up with the external signal within the first 0.1~$\mu$s. The leading distance has a minor oscillation in the initial stage and becomes steady after 0.5~$\mu$s. If we artificially remove the MTJs from the network, i.e., the STD effect does not exist and $p\equiv1$, the firing rate of the 1D CANN will be as plotted by the green dashed line in Fig.~\ref{fig:4}(a)--(e). Then, the bump of the firing rate always has a finite delay relative to the external signal. We note that the external signal moves only towards the right (counterclockwise) in Fig.~\ref{fig:4}. If the signal abruptly changes its moving direction, the response of the CANN can follow the updated moving direction and quickly achieve anticipative tracking again. The dynamic process is explicitly shown in the Supplemental Material \cite{SM}.

\begin{figure}[t]
\includegraphics[width=0.9\columnwidth]{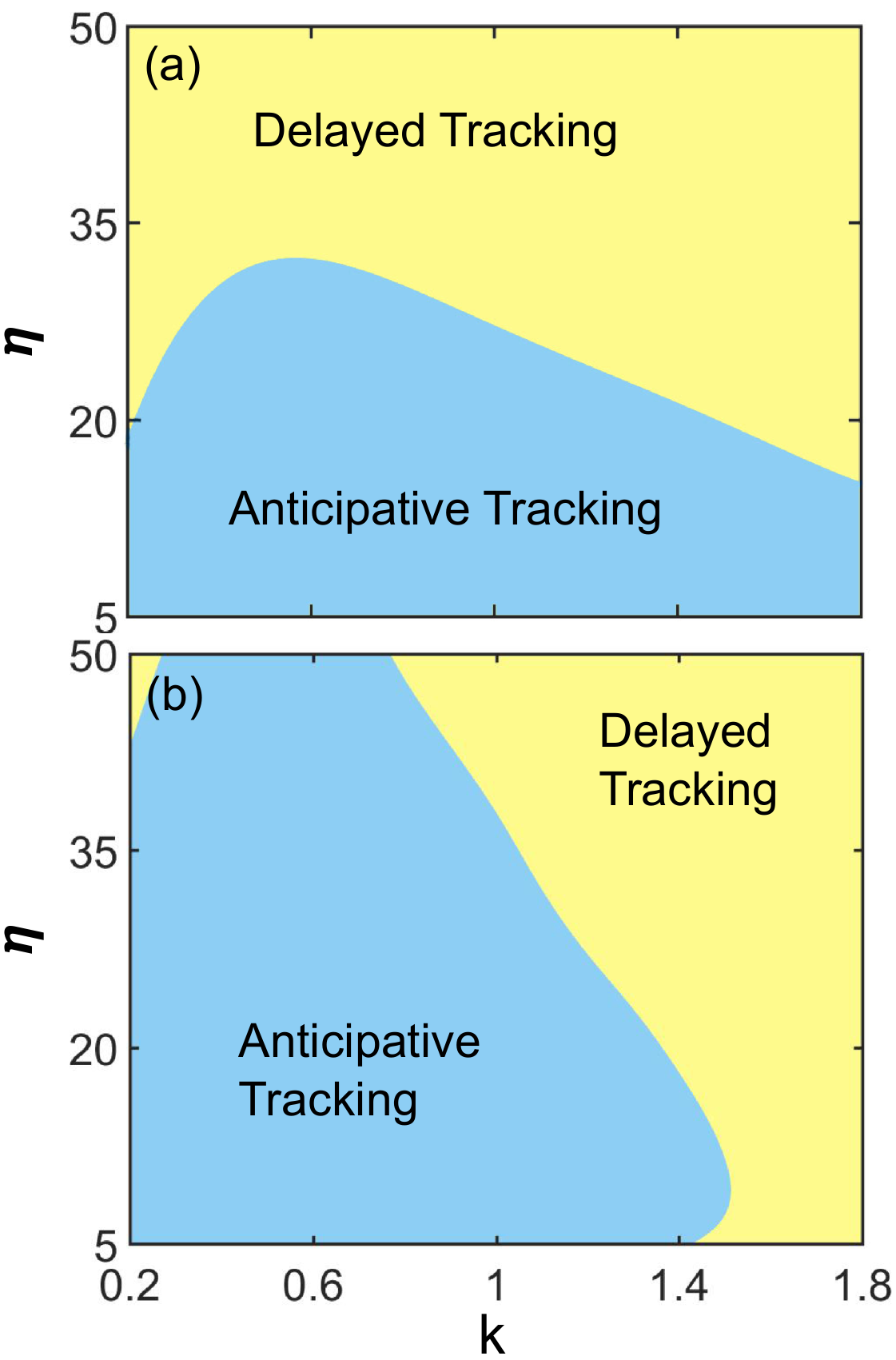}
\caption{The phase diagram of the 1D CANN. The external signal moves with a constant angular velocity $\omega_{\rm ext}=0.005$~rad/ns (a) and $\omega_{\rm ext}=0.01$~rad/ns (b). The graph shows the tracking performance of the network achieved by varying the STD strength $\eta$ and the inhibition strength $k$. It is examined by comparing the central positions of the firing rate bump and the external signal at $t=80$~ns. }\label{fig:5}
\end{figure}
The global inhibition strength $k$ and the STD strength $\eta$ are two key parameters that influence the performance of the 1D CANN in addition to the STD time scale $\tau$. In the above computation, we use fixed values $k=1$ and $\eta=0.8$. We now examine the prediction functionality of the 1D CANN for a constant $\omega_{\rm ext}=0.005$~rad/ns and vary the two parameters in a broad range from $0.2\le k\le 1.8$ to $5\le\eta\le50$. The phase diagram of the CANN is shown in Fig.~\ref{fig:5}(a). For every $k$, there is a boundary for $\eta$, below which anticipative tracking can be realized. This is because a larger $\eta$ in Eq.~\eqref{eq:stdj} results in a larger current density $j$ and hence a stronger STD with an immediate saturation of the synaptic weight. Thus the dynamical synapses loss their functionality of the short-term plasticity. On the other hand, if $\eta$ is extremely small, the current density is then nearly constant corresponding to the vanishing STD. So there is also a lower boundary for $\eta$, which is not shown in this figure. The boundary of the delayed and anticipative tracking exhibits a weak, nonmonotonic dependence on $k$, indicating the presence of a wide range of parameters to realize anticipative tracking.

Figure~\ref{fig:5}(b) shows the phase diagram of the same 1D CANN for tracking a moving object with a larger speed $\omega_{\rm ext}=0.01$~rad/ns. The range of $\eta$ for anticipative tracking increases because for a very mobile signal in the CANN, a weaker driving force can also sustain the spontaneous traveling wave \cite{Wu2005}. For the relatively large $k$, both the upper and lower boundaries of $\eta$ can be seen in the phase diagram. Moreover, a smaller $k$ is needed for a faster object. This is partly because the inhibition parameter $k$ is to avoid the divergence of the network. A faster object results in a less local stimulus for every neuron and the smaller inhibition strength can be used to allow the anticipative tracking. On the contrary, a large $k$ leads to a quick dissipation of signals in the CANN, which can not sustain the spontaneous traveling waves used for anticipative tracking.

%%%%%%%%10%%%%%%%%20%%%%%%%%30%%%%%%%%40%%%%%%%%50%%%%%%%%60%%%%%%%%70%%%%%%%%80

\subsection{Tracking a moving ball in a video}

\begin{figure}[t]
\includegraphics[width=\columnwidth]{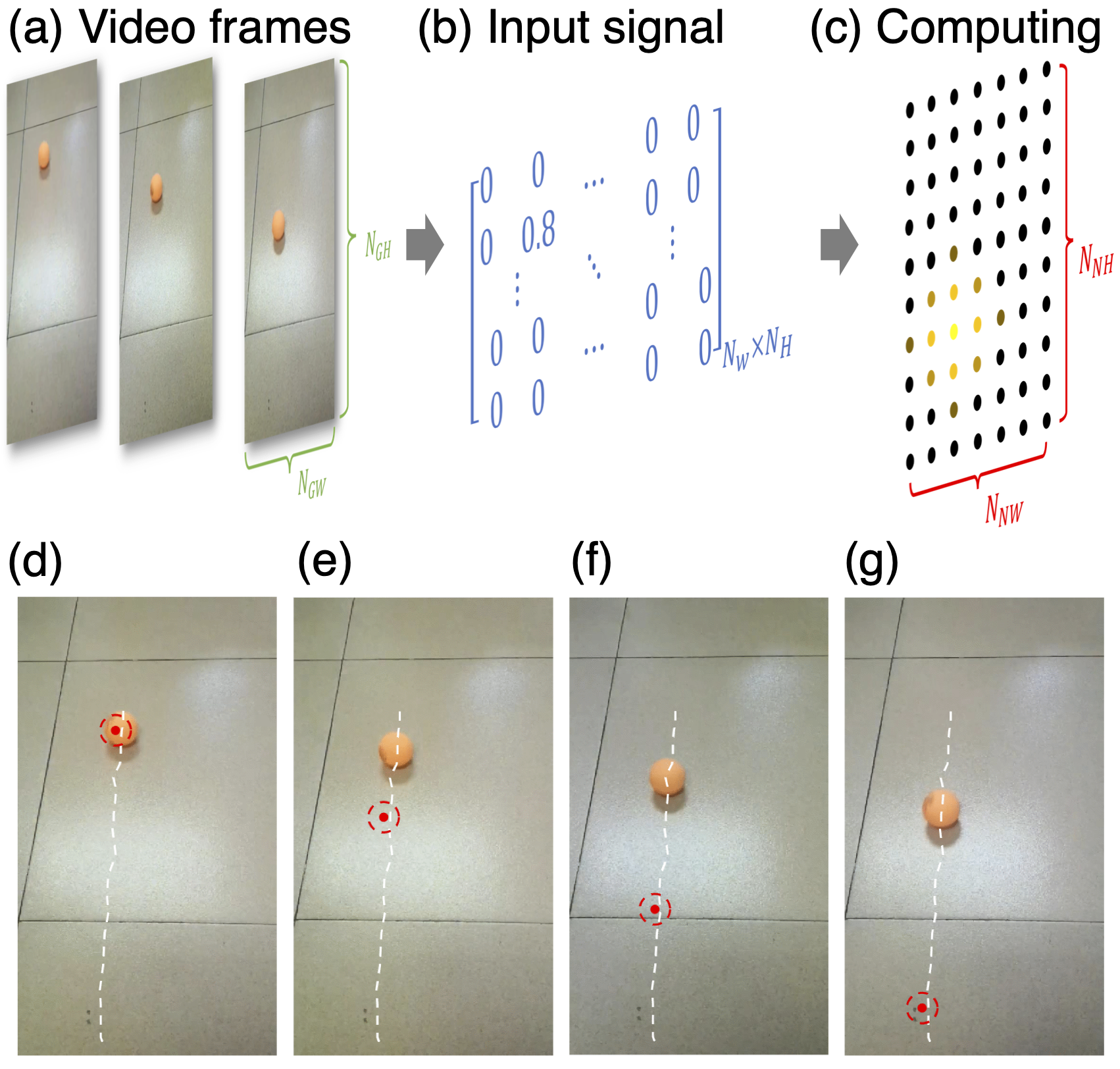}
\caption{(a)--(c) Schematic illustration of the computational process of the 2D CANN. The $544 \times 960$-pixel frames of the recorded video are first converted into an external input matrix by integrating every group of $32 \times 32$ pixels into one matrix element. Every element containing the orange rolling ball is set as 0.8, while the others, as 0. The dimensions of the 2D CANN are $N_{NW} \times N_{NH} = 27 \times 40$, including additional elements on each side of the boundary to avoid a technical problem due to the broken translational symmetry. After the computation using the 2D CANN, the location of the maximum firing-rate value is determined using the cubic spline interpolation. This location represents the predicted moving direction of the rolling ball. (d)--(g) Illustration of the prediction. The red circles are the real-time prediction at the given frame, which closely follows the real trajectory of the rolling ball indicated by the white dashed line.}\label{fig:6}
\end{figure}

The 1D CANN presented above is only a simplified model for demonstrating how a CANN combined with STD works in the prediction of a moving object. Here, we switch to a realistic case, where a moving object is recorded in a video. A 2D CANN is then considered, with arrays of neurons associated with MTJs in a square lattice. The particular process is schematically shown in Fig.~\ref{fig:6}(a)--(c). A video is taken by a cell phone, in which a table tennis ball is rolling on the ground and 60 frames are extracted to examine the prediction of the ball motion using the 2D CANN. The resolution of the original video is $N_{GW} \times N_{GH}=544\times960$ pixels, as shown in Fig.~\ref{fig:6}(a). We first integrate every $32 \times 32$ pixels of the video as one element of the external signal, and every frame is therefore converted into an $N_{W} \times N_{H}=17\times30$ matrix; see Fig.~\ref{fig:6}(b). To avoid artifacts due to the boundary, we apply 5 additional rows or columns of neurons to every side of the frame until the number of neurons in the 2D CANN $N_{NW} \times N_{NH} = 27 \times 40$. The constant connection is given in Eq.~\eqref{eq:const}, with $a=35$ and $b=0.24$, and the global inhibition strength $k$ is set as 0.11. The dynamical synapses are realized by the same MTJs used in the 1D CANN and for $\eta=5$.

Without involving complex signal processing, we simply define the external signal using its color in the corresponding pixels. Specifically, the external input is set as $0.8$ for the position of the orange ball and $0$ otherwise. Then, the input for every frame is transferred into the 2D CANN for a period of 4~ns, which is an unrealistically fast speed of the moving target. This is done to match the short time scale of the STD using the particular MTJs because of the limited computational power for the micromagnetic simulations in practice. The predicted moving direction is determined by the maximum firing-rate location of the 1128 neurons, which is marked by a red circle in every frame. We take a few frames as examples and plot the red circles in Fig.~\ref{fig:6}(d)--(g). These circles closely follow the white dashed line, which denotes the real rolling trajectory of the table tennis ball. The successful tracking result is shown in an animation in graphics interchange format in the Supplemental Material~\cite{SM}. Note that the distance of the predicted position ahead of the real-time position is not a constant, but increases as a function of time instead. Making the quantitative prediction in time and/or position is worth investigating in the future.

\section{Discussions}

Tracking and predicting a moving object using CANNs does not require training in advance and prediction can occur for a broad range of speeds with the appropriate network parameters, such as the decay time $\tau$ of the STD, the STD strength $\eta$ and the inhibition strength $k$. In principle, $\eta$ and $k$ can be tunable in practice indicating the flexibility of the tracking device. Another advantage is that tracking can be achieved without the prior identification of the object in contrast to other algorithms. Therefore,  the main process required for a tracking device is converting the image captured by high-speed cameras into electrical signals and passing the signals to the CANN. This scheme saves lots of computational effort used for image recognition using, e.g., a convolutional neural network and significantly increases the efficiency. Supposing a high-speed camera captures one million of frames per second, one has the time interval 1 $\mu$s to deal with every frame. The magnetization dynamics is usually at the time scale of 10--100 ns, so it won't be a bottleneck for the process. A quantitative estimation of the time consumption requires the detailed information of the whole circuit and is beyond the scope of the present work.

The network can capture multiple moving elements, as illustrated in the Supplemental Material \cite{SM} with a simple example. Here we define a 2D plane, where two continuous signals moving around. We assume that the time step between two frames is 4~ns. One signal has a circular trajectory with the radius 2 and $\omega=0.9$~rad/ns while the other follows a $3\times3$ square with the speed 0.03~ns$^{-1}$. We then set up a CANN with 21$\times$21 neurons with dynamical synapses. The corresponding parameters are $a=0.5$, $b=0.656$, $\eta=0.1$ and $k=5$. The firing rates of the neurons naturally have two local maxima demonstrating that the two moving objects are both well tracked in the simulation. Nevertheless, since the objects are not identified, distinguish different objects need additional effort. Finally, we would like to emphasize that the noisy background does not affect the tracking since CANNs are sensitive to the continuous motion instead of random noise \cite{Wu2005}. 

In the above 1D and 2D CANNs, the parameters used in the network, such as $k$, $\eta$, $a$, and $b$, are quite different because they are related to the structure and connections of the network. For example, the neurons in the 2D CANN have more neighbors compared with the 1D case and a larger $a$ is hence used in the 2D network to avoid receiving too many signals from the connected neurons. Then the fixed synaptic weights $J_{ij}$ and $r_i$ in the 2D CANN have a relatively low values. This is reason why we have used a larger $\eta$ to ensure the STD effect. Because the overall signal strength is relatively low, a small inhibition parameter $k$ is enough to keep the 2D CANN working. Another reason for the different $a$ we choose is because the distance is 1 between the nearest neighboring neurons in the 2D CANN while that in the 1D case is 0.1$\pi$. For a hardware implementation, if $\eta$, $a$ and $b$ are fixed, one still can choose different current density $j_{\max}$ and $j_{\min}$ to optimize the performance of CANNs.

%%%%%%%%10%%%%%%%%20%%%%%%%%30%%%%%%%%40%%%%%%%%50%%%%%%%%60%%%%%%%%70%%%%%%%%80
\section{Conclusions}

We have demonstrated using a micromagnetic simulation that magnetic tunnel junctions can be used as the hardware of dynamic synapses, which have short-term synaptic plasticity. Continuous attractor neural networks with MTJ-based dynamical synapses are shown to have the ability to anticipatively track high-speed moving objects. A 1D CANN consisting of twenty neuron is used to illustrate in detail how the dynamical synapses are implemented and contribute to the modulation of the firing rates of neurons. The latter is used to predict the position of a moving signal. In a realistic example, we show that a rolling table tennis ball in a video can be tracked and its motion predicted using a 2D CANN including $27\times40$ neurons, where every neuron is associated with an MTJ. Although the tracking task can in principle be accomplished using the CMOS circuits, the hardware implementation of the dynamical synapses with MTJs, utilizing their inherent dynamical properties, is straightforward and suitable for a portable, energy-saving terminals.

The demonstration of the MTJ-based dynamical synapses significantly expands the promising and powerful functionality of spintronic devices for neuromorphic computing \cite{Grollier16,Zhang2020} to process high-speed dynamical information. Moreover, the simple implementation with no need for training makes this proposal very attractive in application. In addition, the principles in this work can be directly applied in the hardware implementation of neuromorphic chips using resistive, ferroelectric, 2D van der Waals and other materials.

%%%%%%%%10%%%%%%%%20%%%%%%%%30%%%%%%%%40%%%%%%%%50%%%%%%%%60%%%%%%%%70%%%%%%%%80
\begin{acknowledgments}
This work was financially supported by the National Natural Science Foundation of China (Grants No. 11734004, No. 61774018, and No. 31771146), the Recruitment Program of Global Youth Experts, and the Fundamental Research Funds for the Central Universities (Grants No. 2018EYT03 and No. 2018STUD03). Y.Y.M. acknowledges financial support from the Beijing Nova Program (Grant No. Z181100006218118). The computation was supported by the Super Computing Center of Beijing Normal University.
\end{acknowledgments}

\end{document}